\documentclass[mathleft]{an}
\usepackage{graphicx}
\usepackage{times}
\usepackage{amsmath}
\usepackage{natbib}
\bibpunct{(}{)}{;}{a}{}{,}

\newcommand\cm{\,\rm cm}

\newcommand\erg{\,\rm erg}
\newcommand\K{\,\rm K}
\newcommand\Myr{\,\rm Myr}

\newcommand\kms{\,\rm km\,s^{-1}}

\newcommand\pc{\,\rm\,pc}
\newcommand\kpc{\,\rm kpc}
\newcommand\vis{\,\rm cm^{2}s^{-1}}

\newcommand\tms{\!\times\!}
\newcommand\cdt{\!\cdot\!}

\newcommand\xx{\hat{{\mathbf x}}}
\newcommand\yy{\hat{{\mathbf y}}}
\newcommand\zz{\hat{{\mathbf z}}}

\newcommand\U{\mathbf{u}}
\newcommand\mU{\bar{\mathbf{u}}}
\newcommand\B{\mathbf{B}}
\newcommand\mB{\bar{\mathbf{B}}}
\newcommand\Tf{\mathcal{B}}
\newcommand\mTf{\bar{\mathcal{B}}}

\newcommand\EMF{\mathcal{E}}
\newcommand\SN{{\rm SN}}

\DeclareMathAlphabet{\mathbf}{OML}{cmm}{b}{it}

\begin{document}


\Pagespan{1}{}
\Yearpublication{2008}%
\Yearsubmission{2007}%
\Month{1}%
\Volume{999}%
\Issue{88}%

\title{Dynamo coefficients from local simulations of the turbulent {ISM}}

\author{O. Gressel\thanks{Corresponding author:{ogressel@aip.de}}, 
        U. Ziegler, D. Elstner \and G. R\"{u}diger}
\institute{Astrophysikalisches Institut Potsdam, 
           An der Sternwarte 16, 14482 Potsdam, Germany}

\titlerunning{Local simulations of the turbulent ISM}
\authorrunning{O. Gressel, U. Ziegler, D. Elstner \& G. R\"{u}diger}

\received{--}
\accepted{--}
\publonline{--}


\abstract{%
Observations in polarized emission reveal the existence of large-scale
coherent magnetic fields in a wide range of spiral galaxies. Radio-polarization
data show that these fields are strongly inclined towards the radial
direction, with pitch angles up to $35\degr$ and thus cannot be explained by
differential rotation alone. Global dynamo models describe the generation of
the radial magnetic field from the underlying turbulence via the so called
$\alpha$-effect. However, these global models still rely on crude assumptions 
about the small-scale turbulence. To overcome these restrictions we perform 
fully dynamical MHD simulations of interstellar turbulence driven by supernova
explosions. From our simulations we extract profiles of the contributing
diagonal elements of the dynamo $\alpha$-tensor as functions of galactic
height. We also measure the coefficients describing vertical pumping and find
that the ratio $\hat{\gamma}$ between these two effects has been overestimated
in earlier analytical work, where dynamo action seemed impossible. In
contradiction to these models based on isolated remnants we always find the
pumping to be directed inward. In addition we observe that $\hat{\gamma}$ 
depends on whether clustering in terms of super-bubbles is taken into account.
Finally, we apply a test field method to derive a quantitative measure of the 
turbulent magnetic diffusivity which we determine to be $\sim 2\,\kpc\kms$.}

\maketitle

\section{Introduction} 

The last decade with its advances in observational technology has brought
increasingly detailed maps of galactic magnetic fields. Radio-polarization
data indicate large scale regular fields within the interstellar medium.  For
typical spiral galaxies the ISM is highly turbulent
\citep{2004RvMP...76..125M}. The main drivers of the turbulence are thought to
be winds of massive stars and explosions of supernovae (SNe)
\citep{2005A&A...436..585D,2006ApJ...653.1266J}, as well as (in less active
regions) the magneto-rotational instability (MRI)
\citep*{2004A&A...423L..29D,2005ApJ...629..849P,2007ApJ...663..183P}.
Furthermore, \citet{2004ApJ...605L..33H,2006AN....327..469H} have shown that a
Parker-type instability can be driven by cosmic rays, an idea first advocated
by \cite{1992ApJ...401..137P}. It, however, remains a matter of discussion,
whether cosmic rays play a significant role in the interstellar dynamics
\citep{2006MNRAS.373..643S}.

The generation of a mean magnetic field from turbulent fluctuations can be
explained via the so called $\alpha$-effect which describes the correlations
of the small-scale turbulent velocity and magnetic field giving rise to a mean
electromotive force (EMF). In the case of the ISM there are three
characteristic asymmetries leading to non-vanishing correlations: (i) the axis
of rotation, (ii) the galactic shear gradient, and (iii) the (vertical)
gradient in density and turbulence intensity \citep{1993A&A...269..581R}. The
mutual strength of the different terms puts constraints on the operability of
a dynamo process. The ratio $\hat{\gamma}$ of the diamagnetic pumping over the
$\alpha$-effect has an influence on the efficiency of the dynamo. The strength
of the $\alpha$-effect compared to the shear puts limits on the pitch angle.

Until recently, a direct numerical simulation of the turbulent ISM has been
infeasible. Therefore, early theoretical models like the SOCA approach by
\citet{1993A&A...269..581R} predicted the outcome of ISM-turbulence based on
simplifying assumptions. In an alternative approach
\citet{1992ApJ...391..188F} derived the dynamo effect for isolated supernova
remnants (SNRs) and super-bubbles (SBs). However, these early no-interaction
models arrived at prohibitively high values for $\hat{\gamma}$. To test these
findings first numerical simulations for single SNRs have been performed in 2D
by \citet*{1993A&A...274..757K} and in 3D by \citet*{1996A&A...305..114Z}.
Still the key issue of a dominating turbulent pumping remained.  The situation
was somewhat improved by taking into account stratification \citep[hereafter
FER98]{1998A&A...335..488F} yielding a value $\hat{\gamma} \approx 6$. The
parameter range for $\hat{\gamma}$ permitting dynamo solutions has been
explored by \citet*{1994A&A...286...72S}.

The major limitation to the no-interaction models described above results from
the fact that (even for SBs) the explosions are considered as isolated events
taking place on a uniform background. But this is not the case for the ISM
which due to thermal instability \citep{1965ApJ...142..531F} is a highly
heterogeneous medium. To overcome these limitations we perform direct
numerical simulations of the local ISM and use a test field method
\citep{2005AN....326..245S,2007GAFD...101..81S} to obtain the dynamo
parameters.

\section{Physical model and parameters} 

We simulate the dynamic evolution of the stratified, turbulent ISM utilizing a
3D MHD model including the physical effects described below. The computational 
domain covers a box of $0.8\times 0.8\times 4.0\,\kpc^3$ vertically centered 
around the midplane and representing a local patch of the galactic disk. The
purely vertical gravitational potential is adopted from 
\cite{1989MNRAS.239..605K}.

The equations of resistive MHD are solved in the local shearingbox approach,
i.e., we apply a co-rotating Cartesian coordinate system with $\xx$, $\yy$,
and $\zz$ being the unit vectors along the radial, azimuthal, and vertical
direction. The background shear of the flow is characterized by the parameter
$q=d \ln \Omega/d \ln R$, where the case $q=-1$ corresponds to a flat rotation
profile. Additional source terms $\Gamma_\SN-\rho^2\Lambda(T)+\rho\Gamma(z)$
in the total-energy equation represent the thermal energy input due to
supernovae and optically thin radiative cooling/heating.

\begin{figure*}
    \center\includegraphics[width=1.8\columnwidth]{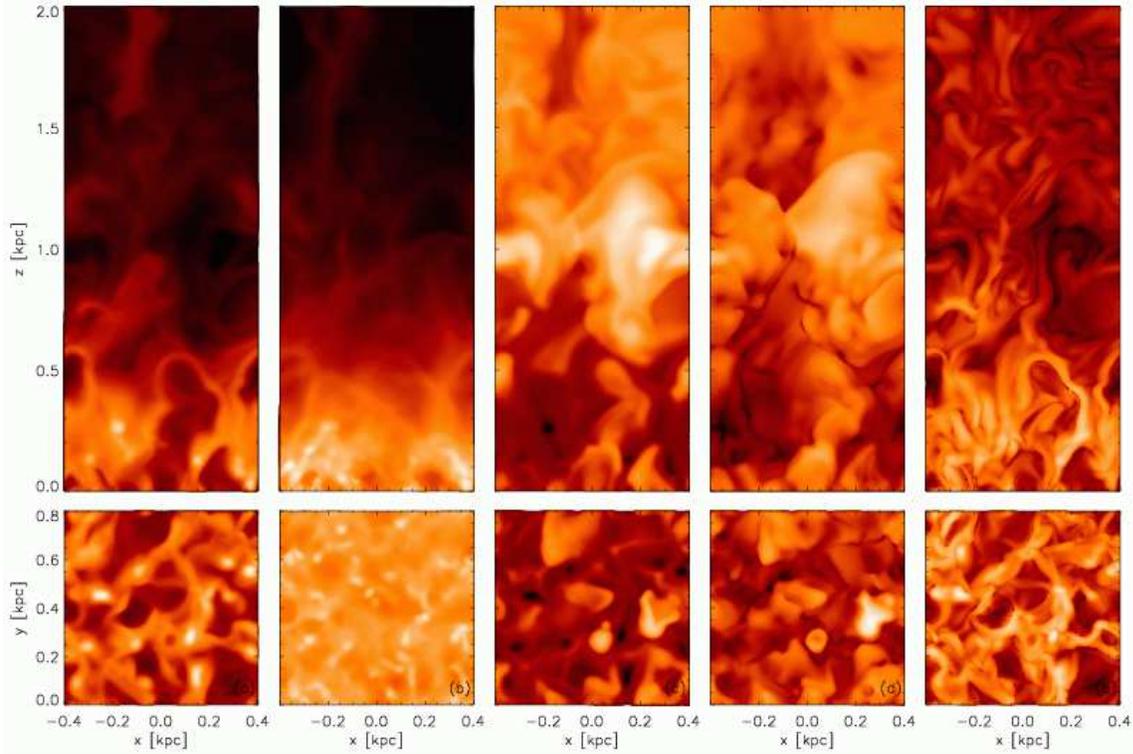}
\caption{Vertical slices of the top half of our box (upper panels),
  and horizontal slices through the midplane (lower panels) after
  $t=112\Myr$. Quantities shown are: (a) number density $[\cm^{-3}]$,
  (b) column density $[\cm^{-2}]$, (c) temperature $[\K]$, (d) velocity 
  dispersion $[\kms ]$, and (e) magnetic field strength $[\mu G]$. 
  The logarithmic grey scales extend over ranges $[-4.82, 1.10]$, 
  $[17.34,21.55]$, $[ 2.06, 7.20]$, $[-0.15, 2.77]$, and $[-4.16, 0.34]$
  for the corresponding variables.}
\label{fig:slices}
\end{figure*}

\subsection{Supernova driving}

We model supernova explosions as local injections of thermal energy resulting in
turbulence at mildly supersonic Mach numbers. To assure convergence of the
emerging supernova remnants numerical solutions with increasing spatial
resolution $\Delta s$ have been tested against the analytical description by
\citet*{1988ApJ...334..252C}. The results agree well and correspond to the
findings reported in \citet{2005ApJ...626..864M}.

SN-events are exponentially distributed in the vertical direction with
scale heights of $325\pc$ for type-I and $90\pc$ for type-II SNe. For the 
models presented in this paper we use 3/4 the galactic frequencies which are 
$\sigma_{\rm I}=4\Myr^{-1}\kpc^{-2}$ respectively 
$\sigma_{\rm II}=30\Myr^{-1}\kpc^{-2}$. The associated explosion energies are
$10^{51}$ and $1.14\tms10^{51}\erg$ \citep{2001RvMP...73.1031F}.

Within our model we make an important distinction between type-I and type-II
SNe. The latter are spatially clustered by the (artificial) constraint that
the density at the explosion site must be above average (with respect to a
horizontal slab) while the former are spatially uncorrelated
\citep{1999ApJ...514L..99K}. We use this simple prescription as a proxy for a
more self-consistent treatment of the clustering \citep{2005A&A...436..585D} and
find a fraction of clustered events comparable to observations
\citep{2001RvMP...73.1031F}. The reference simulation without SBs reveals that 
the general morphology is affected quite strongly by the clustering indicating
the importance of this effect. 

\subsection{Radiative cooling and diffuse heating}

\begin{table}
\center\begin{tabular}{|l||c|c|c|c|c|c|c|}\hline 
& dom. $[\kpc]$& grid& SNe& cl.& q& $\beta_{\rm P}$
\tabularnewline\hline\hline 
STD& $0.8,\pm$2.00& $96^2\tms480$& I+II& y& 0& 2000
\tabularnewline\hline 
NCL& $0.8,\pm$2.00& $96^2\tms480$& I+II& n& 0& 2000
\tabularnewline\hline 
SHR& $0.8,\pm$2.13& $96^2\tms512$& I+II& y&$\!$-1& 2000
\tabularnewline\hline 
SN2& $0.8,\pm$2.00& $96^2\tms480$& II&   y& 0& 2000
\tabularnewline\hline 
KIN& $0.8,\pm$2.00& $96^2\tms480$& I+II& y& 0& $\infty$
\tabularnewline\hline
\end{tabular}
\caption{Overview of conducted models. Clustering ('cl.') applies to
  type-II SNe only. SN-rates are 3/4 the galactic values.\label{tab:models}}
\end{table}

We treat the interstellar medium as an optically thin plasma and prescribe the
coupling to the radiation field via a piecewise power law of the form:
$\Lambda(T) = \Lambda_i\,T^{\beta_i}$, for $T_i \le T < T_{i+1}$. The
parameters used are essentially a combination of the cooling curves given by
\citet*{2002ApJ...577..768S} and \citet{2005MNRAS.356..737S}, where the
coefficients have been slightly modified to make the resulting curve
continuous. In contrast to previous work by \citet{1999ApJ...514L..99K} we
include the thermally unstable regime below $6102\K$ leading to the formation
of a cold ISM phase. To numerically resolve the cooling instability we apply
thermal conduction such that the Field length $\lambda_{\rm F}$ is covered by
4 grid cells \citep[cf. ][]{2004ApJ...602L..25K}. The various effects that
contribute to the diffuse heating of the interstellar plasma are subsumed in a
prescribed heating rate $\Gamma(z)=\Gamma_0\,e^{-z/h}$, with $h=300\pc$ as
discussed in \citet{2006ApJ...653.1266J}.

\subsection{The initial model}\label{sec:strat}
Previous stratified models \citep{1999ApJ...514L..99K,2004A&A...425..899D,
2006ApJ...653.1266J,2007ApJ...663..183P} all start from an isothermal initial
state at a prescribed temperature. The main drawback of this is that the
isothermal stratification is not in radiative equilibrium and the disk will
instantaneously collapse until the dynamic pressure from SN- or MRI-turbulence
will balance this process. To avoid this undesired behavior we propose a more
sophisticated initial model where the vertical profiles of density and
pressure are numerically integrated to be in combined hydrostatic and
radiative equilibrium. The computed profiles are considerably flatter than the
isothermal ones while the temperature varies by a factor of about five.

For our fiducial model we choose an initial midplane density $\rho_0$
corresponding to one particle per $\cm^3$ which results in an equilibrium
pressure of $p_0/k_B=6000\K\cm^{-3}$. We use a mean molecular weight of
$\bar{\mu}=0.6$ assuming a fully ionized plasma of cosmic abundance. For the
current study the rotation rate is fixed at $\Omega=100\kms\,\kpc^{-1}$. As in
\citet*{2007ApJ...654..945B} we scale the dynamic viscosity coefficient with
the density, i.e., we use a constant kinematic viscosity of
$\nu=5\tms10^{24}\,\vis$. To obtain a constant Prandtl number of ${\rm Pr}=4.2$ 
we apply the same scaling to the thermal conduction coefficient $\kappa$.

The magnetic Prandtl number ${\rm Pm}=\nu/\eta$ is thought to be very high for
the ISM. With the limited dynamic range of our simulations we are, however,
restricted to values close to unity. For practical purposes we choose 
${\rm Pm}=2.5$ equivalent $\eta=2\tms10^{24}\,\vis$ which is still two orders of
magnitude smaller than the expected turbulent diffusivities.

The vertical boundary conditions are of the outflow type with the magnetic
field extrapolated according to the solenoidal constraint. Due to the
(sheared-) periodic boundary conditions in the x- and y-direction the vertical
magnetic field is ideally conserved for any horizontal slab. This means that
there are two distinct classes of models including or excluding vertical
magnetic flux. For simplicity we focus on the latter case and start from an
initially toroidal plus radial field with a pitch angle of $-6\degr$.  The
field is scaled with $\sqrt{p/p_0}$ to yield a constant plasma parameter
$\beta_{\rm P}=2p/B^2 \approx 2000$ throughout the disk. The resulting
Alfv{\'e}n velocity for the initial model is $0.3-0.6\kms$ whereas the sound
speed ranges from $10-20\kms$. The particular choice of parameters for the
conducted models is summarized in Table~\ref{tab:models}.

\section{Numerical methods} 

For our computations we make use of the newly developed version 3 of the NIRVANA
code \citep{2004JCoPh.196..393Z,2005CoPhC.170..153Z} which is a general purpose 
MHD fluid tool employing the technique of adaptive mesh refinement (AMR). For
the simulations presented here this feature is, however, not being employed.
We extended the standard shearingbox model to be compatible with the
conservative numerical scheme and apply flux-matching at the sheared
interfaces to improve the conservation properties \citep{2007CoPhC.176..652G}.

\subsection{Dynamo test fields}

To find a closure for the mean field equations one strives to parameterize 
$\EMF=\overline{\U'\tms \B'}$ with respect to averaged quantities. Here we 
adopt the standard description where $\EMF$ depends on the mean field and 
its gradients. When using spatial averages along horizontal slabs  
\citet{2002GApFD..96..319B} showed that the resistivity tensor
can be reduced and one yields:
\begin{equation}
  \EMF_i = \alpha_{ij} \bar{B}_j 
         - \tilde{\eta}_{ij}\varepsilon_{jkl}\partial_k \bar{B}_l\,,
  \quad i,j \in \left\{R,\phi\right\}, k=z\,.
  \label{eq:param}
\end{equation}
While the diagonal elements of ${\boldsymbol \alpha}$ quantify 
the dynamo process, its anti-symmetric off-diagonal elements constitute 
the vertical pumping effect described by the parameter 
$\gamma_z = 0.5\,(\alpha_{\phi R}\!-\!\alpha_{R\phi})$.
Similarly the diagonal elements of ${\tilde{\boldsymbol\eta}}$ are interpreted
as turbulent resistivity $\eta_t$, while its off-diagonal components can lead to
$\mathbf{\Omega\tms J}$-type dynamo effects 
\citep{2004maun.book.....R}. In total we have 4+4 unknowns that we wish to
determine from equation~(\ref{eq:param}).

To avoid complications with the inversion of this tensorial equation, we apply
the test field approach proposed 
by \citet{2005AN....326..245S,2007GAFD...101..81S}. The method has also
recently been adopted to the shearingbox case by \citet{2005AN....326..787B}. 
Earlier approaches \citep{2002GApFD..96..319B,2005mpge.conf..171K} were based 
on least square fit methods. The major drawback with these was that in regions
where $\mB$ or $\nabla \mB$ vanishes the inversion becomes singular. This can
be circumvented by solving equation~(\ref{eq:param}) for fixed test fields 
$\mTf_{(\nu)}$ with simple, well behaved geometry and gradients. To obtain the
related EMFs one has to evolve an extra (passive) induction equation for the
associated fluctuations:
\begin{eqnarray}
  \partial_t \Tf'_{(\nu)} & = & \nabla \times [\ 
    \U'\tms\mTf_{(\nu)} + (\mU\!+\!q\Omega x\yy)\tms\Tf'_{(\nu)} \nonumber\\&\,&
    - \overline{\U'\tms\Tf'}\!_{(\nu)} + \U'\tms\Tf'_{(\nu)} 
    - \eta\nabla\tms\Tf'_{(\nu)}\;] \nonumber\,,\\
  \nabla\cdt\Tf'_{(\nu)} & = & 0\,.\label{eq:testfields}
\end{eqnarray}
We implemented these additional equations within NIRVANA employing the
constrained transport paradigm to exactly satisfy the solenoidal constraint. 
The actual method uses up-winding to guarantee stability while second order in
space is attained via piecewise linear reconstruction. For this we apply the 
same slope limiter as in the actual code. Our procedure is very similar to the 
methods described in \citet*{2006JCoPh.218...44T}.

For the particular choice of the four test fields $\mTf_{(\nu)}$ we use the 
ones from \citet{2005AN....326..787B}, i.e., the lowest Fourier modes in the
vertical direction. For each of the test fields we compute the corresponding
mean electromotive force
\begin{equation}
  \bar{\EMF}^{(\nu)} = \overline{\U'\tms\Tf'}\!_{(\nu)}\,.
\end{equation}
The dynamo coefficients can then be computed via equation~(\ref{eq:param})
to which the solution can be compactly written as
\begin{equation}
\left(\begin{array}{c}
\alpha_{ij}\\
k_1 \eta_{ij3}
\end{array}\right) = 
\left(\begin{array}{cc}
\,\,\cos(k_1\!z) & \sin(k_1\!z) \\ \!\!-\sin(k_1\!z) & \cos(k_1\!z)
\end{array}\right)
\left(\begin{array}{c}
\bar{\EMF}_i^{(2j\!-\!2)} \\
\bar{\EMF}_i^{(2j\!-\!1)} 
\end{array}\right)\,,\label{eq:tf_solution}
\end{equation}
with $i,j\in\{1,2\}$. In contrast to the least square fit method
equation~(\ref{eq:tf_solution}) can be directly computed for each $z$,
yielding vertical profiles for the dynamo parameters.

\section{Results and discussion} 

Since our vertical stratification is in radiative equilibrium, we do not 
observe an initial collapse of the disk. Turbulence builds up smoothly and
after 100Myr reaches a quasi steady state. Fig.~\ref{fig:slices} shows
vertical and horizontal slices through the simulation box at $t=112\Myr$. Most
of the material is contained in cold clumps forming a $150-200\pc$ wide
disk. Close to the midplane the network of clumps and filaments is permeated
by strong shocks from the SNe which are continously creating hot
cavities. Looking at the lower panels (a) and (e) of Fig.~\ref{fig:slices}
one can see that for the region around the midplane there exists a significant
correlation between the density and the magnetic field amplitude. The inferred
slope of the correlation is somewhat steeper than the Chandrasekhar-Fermi
value of $0.5$ but roughly consistent with $|B|\sim\rho^{2/3}$ as indicative
of compressional amplification.

While single SNRs are largely confined to the midplane, super-bubbles break out
of the central disk and drive moderate vertical flows. Their dense shells that
are further compressed by shocks will form clouds that can efficiently cool
and will in turn rain back into the gravitational potential thus forming what is
termed the galactic fountain \citep{1980ApJ...236..577B}. The Mach number of
the flow is mildly supersonic below $1\,\kpc$ and becomes subsonic for the outer
parts.

If we turn off the SN-clustering the morphology changes quite drastically.
Instead of well confined SBs we see more disrupted features and chimney-like
structures that channel strong vertical outflows. The velocity dispersion in
the hot phase is twice as high as in the clustered case. These differences
demonstrate the importance of a proper modeling of clustered explosions.

\subsection{Thermal distribution and velocity dispersions}

\begin{figure}\begin{center}
  \includegraphics[width=\columnwidth]{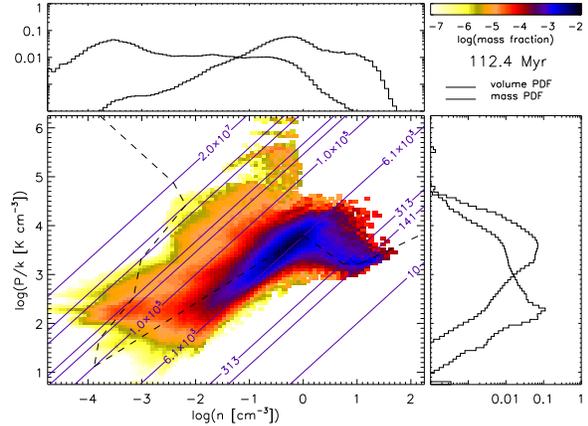}
\caption{Phase space distribution of the SN-heated plasma. Adjacent 
   plots show the mass- (thick) and volume- (thin line) distribution of 
  density (upper panel) and pressure (right panel). We also plot isothermal
  contours (labeled in$\K$), and the equilibrium cooling curve (dashed).}
\label{fig:phs_spc}
\end{center}\end{figure}

In Fig.~\ref{fig:phs_spc} we show the distribution of the SN-heated
plasma as a function of density and thermal pressure at a time $t=112\Myr$. 
The two stable branches of the equilibrium curve are densely populated but
there also exists a considerable amount of gas in the radiatively unstable 
regimes.

Averaged velocity dispersions for the ISM phases are $5\kms$ (cold), 
$11\kms$ (cool), $25\kms$ (warm), and $40-60\kms$ (hot). While the values 
for the latter are consistent with the findings of \citet{2005A&A...436..585D},
for the cold phases we fall short by a factor of two. We consider this a
resolution issue but it might also be related to the different form of the
driving. As a function of galactic height the time averaged rms-velocity rises
steeply from its midplane value of $20\kms$ to its peak value of $55\kms$
which is reached at about $1\,\kpc$ and then falls off to about $30\kms$ at
$2\,\kpc$. Assuming that diamagnetic pumping follows the inverse gradient in 
turbulence intensity this implies an inward transport of magnetic flux for the
central part of the disk.

\subsection{Dynamo parameters}

\begin{figure}\begin{center}
  \includegraphics[width=0.9\columnwidth]{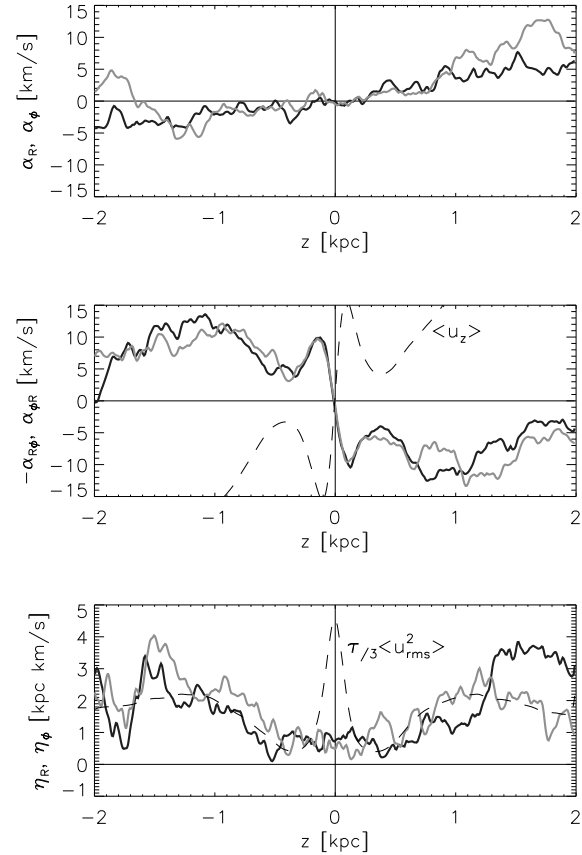}
\caption{Dynamo $\alpha$- and $\eta$-coefficients for the STD  model. Quantities
 indicated by the ordinate labels are plotted in 
 dark ($\alpha_{RR},\dots$) respectively light ($\alpha_{\phi\phi},\dots$)
 colors. Shaded areas indicate the rms-fluctuation when applying the method to
 4 temporal sub-intervals.}
\label{fig:alpha_prof}
\end{center}\end{figure}

In Fig.~\ref{fig:alpha_prof} we plot the main $\alpha$- and
$\eta$-coefficients for our standard run (mod. STD). To our knowledge this is
the first time such profiles have been obtained from direct simulations of
SN-turbulence. The corresponding vertically averaged amplitudes are listed in
Table~\ref{tab:results} -- for comparison we also list the peak values from
FER98 (where $\eta_t$ is distinguished into horizontal and vertical part).
Dynamo numbers $C_{\alpha}=\alpha H/\eta_t$ and $C_{\Omega}=\Omega H^2/\eta_t$
are computed with $H=0.8\,\kpc$ and $\eta_t$ values from the inner part of the
disk. In accordance with analytical estimations based on these numbers, the
runs without shear are still sub-critical, i.e., no field-amplification is
observed. The $\alpha^2\Omega$-dynamo, however, does exponentially grow at a
time scale of $\sim 250\Myr$. 

During the course of the present simulations we do not yet reach the
equipartition field strength. This implies that the values for the dynamo
parameters (as well as the pitch angle) are representative of the unquenched
regime.
Our values are still affected by fluctuations and a longer time-base is needed
for a more accurate determination. However, we find some interesting trends in
the data at hand: The antisymmetric part $\gamma_z$ of the $\alpha$-tensor is 
negative (positive) in the northern (southern) hemisphere, i.e., pumping is 
directed inward. This is contrary to the predictions by Ferri{\`e}re for the 
case of non-interacting SNRs. In our simulations the inward pumping
is opposed by an outward advection of the field with the mean flow. As can be
seen from the middle panel of Fig.~\ref{fig:alpha_prof} both terms have
about the same magnitude for $z < 1\,\kpc$ while in the corona the wind is
dominating. The balance in the inner region could be seen as an indication
that the effects of turbulent pumping and advection are naturally linked. In
consequence vertical transport processes might be of lesser importance than
formerly believed.

The importance of modeling spatially coherent SBs can be seen from comparing
the first two rows of Table~\ref{tab:results}. In the case without clustered
SNe (mod. NCL) we observe strong vertical streaming motions reflected in high
values for $\gamma_z$ and $\bar{u}_z$. Also the turbulent diffusivity is high
in the disk midplane which is not the case for the other runs where $\eta_t$
increases with galactic height. From the model with type-II SNe only (mod.
SN2) one can learn that the turbulent diffusivity predominantly arises from
the more broadly distributed type-I SNe -- despite their lower rate by a
factor of eight. The overall level of turbulent diffusion is about
$3$--$6\tms10^{26}\,\vis$. The off-diagonal components of the $\eta$-tensor
are somewhat smaller and cannot yet be determined accurately. Following
\citet{1996ApJ...457..798J} we crudely estimate that the measured amount of
diffusion suffices to damp short wavelength MRI-modes for reasonably high
$\beta_{\rm P}$ -- a definite conclusion, however, requires further
investigations by means of combined direct simulations.

By comparing $\alpha_{\phi\phi}$ with the SOCA results by 
\citet{1993A&A...269..581R} 
we can estimate the coherence time $\tau$ and the Coriolis number
$\Omega^*=2\tau\Omega$ of our turbulent flow. We find a value of
$\tau\approx3.6\Myr$ which is about a factor of three smaller than commonly 
assumed. The even lower value for our model NCL indicates that a higher
coherence time might be achieved by a more realistic prescription for the
modeling of SBs. In general the obtained profiles are consistent with their
SOCA counterparts. Our obtained value for 
$\hat{\gamma}=|\alpha_{\phi R}|/|\alpha_{\phi\phi}|$ agrees with the SOCA
value of $\hat{\gamma}\approx2.5$ in \citet{2004maun.book.....R} and is 
smaller than the value of $\hat{\gamma}\approx6$ in FER98.

In the lower panel of Fig.~\ref{fig:alpha_prof} we overplot the scalar
expression for the turbulent diffusivity which matches the profiles obtained
from the simulations. The peak of the velocity dispersion close to the
midplane is an artifact due to the static distribution of the SNe resulting in
a disruption of the central disk. This issue has meanwhile been resolved.

\begin{table*}\begin{center}%
\begin{tabular}{|l|c|c|c|c|c|c|c|c|c|c|}\hline
&$|\alpha_{RR}|$&$|\alpha_{\phi\phi}|$&$|\gamma_z|$&$\hat{\gamma}$
&\multicolumn{2}{c|}{$\eta_{t}$}&$\tau$&$\Omega^*$&$C_{\alpha}$
\tabularnewline
& $[\kms]$ & $[\kms]$ &$[\kms]$ & &\multicolumn{2}{c|}{$[\kpc\kms]$}
& $[\Myr]$ & & 
\tabularnewline\hline\hline
STD & 3.1 & 3.8 & 7.8 & 2.1 & ~~0.9~~ & ~~2.7~~ & 3.6 & 0.7 & 3.3 
\tabularnewline\hline
NCL & 7.3 & 4.3 & 24.7  & 5.9 & 2.4 & 2.2 & 2.8 & 0.6 & 1.4 
\tabularnewline\hline
SN2 & 0.8 & 0.8 & 1.8 & 2.5 & 0.6 & 1.0 & 3.6 & 0.7 & 1.2 
\tabularnewline\hline
KIN & 3.2 & 2.0 & 8.3 & 4.1 & 0.6 & 1.2 & 3.4 & 0.7 & 2.6 
\tabularnewline\hline\hline
FER98 & 6.0 &  2.6 & 16.0 &  6.2&  \multicolumn{2}{c|}{3 -- 18} & 
 -- & -- & 0.5 
\tabularnewline\hline
\end{tabular}
\end{center}
\caption{Mean values of extracted parameters averaged over $\approx150\Myr$. 
Values for $\eta_t$ apply to the inner ($|z|\le 0.8\,\kpc$) respectively outer 
region of the disk. Coherence time $\tau$ and Coriolis number $\Omega^*$
are estimated from a comparison with SOCA-profiles for $\alpha_{\phi\phi}$.
The coefficients could not be obtained for model SHR.
\label{tab:results}}
\end{table*}

\subsection{Pitch angles}
In our models without shear we observe pitch angles ranging from $+10\degr$ in
the far outer regions to $-45\degr$ near the midplane. This is hardly surprising
since also the radial and azimuthal dynamo effects are found to be of equal
strength in this case. If we include galactic shear with $q=-1$, pitch angles
become solely negative reaching values up to $-30\degr$ as consistent with
observations. The minimum of $-10\degr$ is found at the midplane.
In comparison, \cite{2006AN....327..469H} in their simulations of a cosmic 
ray driven galactic dynamo find the azimuthal field to be dominating by a
factor of ten which corresponds to pitch angles of only $\sim 5\degr$.

\section{Summary} 

We have performed resistive MHD simulations of the differentially rotating,
stratified local interstellar medium. By integrating a radiatively stable
initial stratification we were able to avoid the disk collapse occurring in 
previous isothermal models. We have further demonstrated that SNe can indeed
produce an $\alpha$-effect that is  not dominated by vertical pumping. Our key
findings are as follows:
\begin{itemize}
\item The turbulent velocity dispersion in our box increases with galactic
  height, i.e., the diamagnetic pumping is directed inward.

\item We observe a mean vertical outflow of the same magnitude as the
  turbulent pumping. In consequence this drastically reduces the effective
  vertical transport of the mean magnetic field.
  
\item The ratio $\hat{\gamma}$ of vertical pumping over the $\alpha$-effect 
  is diminished if type-II SNe are modeled as clustered events.

\item The radial pitch angles found in our simulations are consistent with
  observations and suggest an $\alpha^2\Omega$-dynamo.
\end{itemize}

\acknowledgements OG thanks Axel Brandenburg for discussing the test field
method. This work used the NIRVANA code v3.3 developed by Udo Ziegler at the
AIP. All computations were performed on the local sanssouci-cluster. This work
was supported by Deutsche Forschungsgemeinschaft (DFG) under grant Zi-717/2-2.

\end{document}